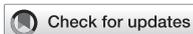

# New developments in econophysics: Option pricing formulas

Moawia Alghalith*

Department of Economics, UWI, St Augustine, St. Augustine, Trinidad and Tobago

We synthesize and discuss some new developments in econophysics. In doing so, we focus on option pricing. We relax the assumptions of constant volatility and interest rate. In doing so, we rely on the square root of the Brownian motion. We also provide simple, closed-form pricing formulas for the American and Bermudan options.

## Introduction

In previous decades, some obstacles existed in econophysics and finance. For example, the free-boundary problem created a serious challenge to pricing American options and related derivatives. Consequently, the previous literature did not offer an explicit, simple formula to price these derivatives (see, for example, Merton [1] and Heston [2]). A similar challenge was evident under the assumption of stochastic volatility or stochastic interest rate.

Furthermore, in theoretical physics and econophysics, some researchers sought the introduction of a new stochastic process. The process is the square root of the Brownian motion. The usefulness of this process was demonstrated by Frasca and Farina [3] and Frasca [4]. In this review, we will briefly discuss the contributions that overcame these obstacles.

## Review

Alghalith [5,6] introduced this PDE for the price of the American call options under the assumption of constant volatility and interest rate

$$C_t + r(SC_S - C) + \frac{1}{2}\sigma^2 S^2 C_{SS} + (1-r)c = 0, C(T, S(T)) = g(S), \quad (1)$$

where c is the consumption at time zero, r is the interest rate, σ is the volatility, S is the price of the underlying asset, g is the payoff, t is time, and C is the price of the option. This can be expressed as

$$C_t + rSC_S + \frac{1}{2}\sigma^2 S^2 C_{SS} - \alpha C = 0, C(T, S(T)) = g(S), \quad (2)$$

where $\alpha \equiv r - \Psi(1 - r)$ and $\Psi$ is a constant. The inclusion of consumption allowed us to circumvent the free-boundary problem. Eq. 2 is a generalized Black-Scholes PDE Black and Scholes [7]; its solution is

$$\begin{aligned} C(t, S) &= e^{\Psi(1-r)(T-t)} \left[ SN(d_1) - e^{-r(T-t)} KN(d_2) \right] \\ &= e^{\Psi(1-r)(T-t)} C_{BS}, \end{aligned} \quad (3)$$

where CBS is the Black-Scholes price of the equivalent European call option, $d_1 = \frac{\ln(S/K)+(r+\sigma^2/2)(T-t)}{\sigma\sqrt{(T-t)}}, d_2 = d_1 - \sigma\sqrt{(T-t)}$ and K is the strike price.

The price of the American put option is given by

$$\begin{aligned} P(t, S) &= e^{\Psi(1-r)(T-t)} \left[ e^{-r(T-t)} KN(-d_2) - SN(-d_1) \right] \\ &= e^{\Psi(1-r)(T-t)} P_{BS}, \end{aligned} \quad (4)$$

where PBS is the Black-Scholes price of the equivalent European put option.

Alghalith [8] showed that the price of the Bermudan put option is given by

$$\begin{aligned} P &= e^{.5\left(e^{r(T-\hat{T})}-1\right)(1-r)T} \left[ e^{-rT} KN(-d_2) - SN(-d_1) \right] \\ &= e^{.5\left(e^{r(T-\hat{T})}-1\right)(1-r)T} P_{BS}, \end{aligned} \quad (5)$$

where $\hat{T}$ is the first possible exercise date. We note that the Bermuda option is a restricted form of the American option.

A limitation of the Black-Scholes model is the assumption of constant volatility. However, the empirical evidence indicates that the volatility changes with time. There is evidence of leptokurtosis and volatility clustering. Similarly, in the real world, the interest rate is not constant.

In order to obtain a (simple) pricing formula under stochastic volatility, Alghalith [9] assumed that the dynamics of the price of the underlying asset are given by

$$dS_u = S_u(\mu du + \beta dW_{1u}) + dS_u \lambda \omega_u \sqrt{dW_{2u}}, \quad (6)$$

where Wu is a Brownian motion λ is a constant, ωu and dBu are independent random variables with zero means; the process $\sqrt{dB_u}$ was introduced by Frasca and Farina [3] and Frasca [4]. Later, Frasca et al [10] formally developed the process of the square root of the Brownian motion. Consequently, the price of the option is

$$C(t, S) = SN(d_1) - e^{-r(T-t)} KN(d_2), \quad (7)$$

where $d_1 = \frac{\ln(S/K)+(r+\beta^2/2)(T-t)}{\sqrt{\beta^2(T-t)}}, d_2 = d_1 - \sqrt{\beta^2(T-t)}$, and β is a constant. It is worth emphasizing that β here is not the volatility of the return rate; it is a component of the volatility.

Alghalith [11] offered a similar formula under the assumption of a stochastic interest rate.

## Conclusion

To conclude, the above-mentioned contributions substantially generalized and extended the Black-Scholes model, while maintaining simple pricing formulas. These contributions are expected to open new paths in econophysics, especially in the area of derivatives. These methods can also be applied to other derivatives, such as the Asian options.

## Data availability statement

The original contributions presented in the study are included in the article/Supplementary Material, further inquiries can be directed to the corresponding author.

## Author contributions

The author confirms being the sole contributor of this work and has approved it for publication.

## Acknowledgments

Author very grateful to Editor SS and the reviewers for their excellent and fast comments.

## Conflict of interest

The author declares that the research was conducted in the absence of any commercial or financial relationships that could be construed as a potential conflict of interest.

## Publisher's note

All claims expressed in this article are solely those of the authors and do not necessarily represent those of their affiliated organizations, or those of the publisher, the editors and the reviewers. Any product that may be evaluated in this article, or claim that may be made by its manufacturer, is not guaranteed or endorsed by the publisher.